\documentclass[12pt]{article} 
\usepackage{graphicx,subfigure,latexsym}
\usepackage{amsmath,amssymb,amsfonts}
\usepackage{bm}

\newcommand{\be}{\begin{equation}}
\newcommand{\ee}{\end{equation}}
\newcommand{\bear}{\begin{eqnarray}}
\newcommand{\eear}{\end{eqnarray}}
\newcommand{\ba}{\begin{array}}
\newcommand{\ea}{\end{array}}

\def \be {\begin{equation}}
\def \ee {\end{equation}}

\def \bes {\begin{subequations}}
\def \ees {\end{subequations}}

\def \<{\langle}
\def \>{\rangle}
\def \+{\dagger}
\def \({\left(}
\def \){\right)}
\def \[{\left[}
\def \]{\right]}

\def \e {\epsilon}

\begin{document}

\begin{titlepage}
\vfill
\begin{flushright}
{\normalsize RBRC-1032}\\
\end{flushright}

\vfill
\begin{center}
{\Large\bf  Spin Polarized Photons and Di-leptons from Axially Charged Plasma}

\vskip 0.3in

\vskip 0.3in
Kiminad A. Mamo$^{1}$\footnote{e-mail: {\tt kabebe2@uic.edu}} and
Ho-Ung Yee$^{1,2}$\footnote{e-mail: {\tt hyee@uic.edu}}
\vskip 0.15in

{\it $^{1}$ Department of Physics, University of Illinois, Chicago, Illinois 60607}\\[0.15in]
{\it $^{2}$ RIKEN-BNL Research Center, Brookhaven National Laboratory,}\\
{\it Upton, New York
11973-5000}\\[0.15in]
{\normalsize  2013}

\end{center}

\vfill

\begin{abstract}

Axial charge in a QCD plasma is P- and CP-odd. We propose and study P- and CP-odd observables in photon and di-lepton emissions from an axially charged QCD plasma, which may provide possible experimental evidences of axial charge fluctuation and triangle anomaly in the plasma created in heavy-ion collisions.
Our observables measure spin alignments of the emitted photons and di-leptons, and are shown to be related to the imaginary part of chiral magnetic conductivity
at finite frequency-momentum, which ultimately arises from the underlying triangle anomaly of the QCD plasma with a finite axial charge density. We present an exemplar computation of these observables in strongly coupled regime using AdS/CFT correspondence.

\end{abstract}

\vfill

\end{titlepage}
\setcounter{footnote}{0}

\baselineskip 18pt \pagebreak
\renewcommand{\thepage}{\arabic{page}}
\pagebreak

\section{Introduction}

Triangle anomaly (or chiral anomaly) of the axial symmetry in QCD with (nearly) massless quarks is
a result of an interesting quantum mechanical interplay between spins, helicities and charges of the fundamental
fermionic constituents of the matter we observe in the Universe.
It dictates that the axial current conservation, which naively holds true in classical limit, is violated quantum mechanically by \footnote{This is the covariant form of anomaly with a single Dirac fermion of charge $e$.},
\be
\partial_\mu J^\mu_A={e^2\over 2\pi^2}\vec E\cdot\vec B\,,
\ee
 in the presence of a P- and CP-odd
environment provided by a non-zero $\vec E\cdot\vec B$.
Its physics consequences are rich, in both low and high temperature/density phases of QCD matter,
and quite recently a lot of interests have been attracted to some of its effects in a high temperature
quark-gluon plasma created in heavy-ion collisions and their possible experimental observations,
which could be one of the direct experimental tests of the fundamental symmetries of QCD.

One such phenomenon is the Chiral Magnetic Effect (CME) \cite{Kharzeev:2007tn,Kharzeev:2007jp,Fukushima:2008xe,Son:2004tq,Metlitski:2005pr} which dictates the existence of an electromagnetic current in the presence of a background magnetic field,
\be
\vec J={e^2\mu_A\over 2\pi^2}\vec B\,,
\ee
where $\mu_A$ is the axial chemical potential. In off-central heavy-ion collisions, the
ultra-relativistic heavy-ion projectiles can create a huge magnetic field which provides an ideal set-up
for CME \cite{Kharzeev:2007jp}, and the axial charges may be created event-by-event either by the glasma color fields in the early stage of collisions or by thermal sphaleron transitions in a later stage \cite{Kharzeev:2007jp,Kharzeev:2001ev,Lappi:2006fp}.
The induced event-by-event charge separation from the CME may lead to
some experimental signatures \cite{Voloshin:2004vk} that indeed seem to be consistent with the observations in RHIC \cite{Abelev:2009ac} and LHC \cite{Selyuzhenkov:2011xq}. However, as the proposed signal is roughly the square of the charge separations in order to avoid
event-averaging to zero, the signal is in fact P-even and may get additional contributions from other background effects unrelated to triangle anomaly \cite{Asakawa:2010bu,Bzdak:2009fc,Wang:2009kd,Pratt:2010zn}, which makes it hard to draw definite conclusions on the CME in heavy-ion collisions.

Another related phenomenon is the Chiral Magnetic Wave (CMW) \cite{Kharzeev:2010gd,Newman:2005hd} which is a gapless sound-like
propagation of chiral (that is, left-handed or right-handed) charges along the direction of the magnetic field. The CMW may lead to a non-zero electric quadrupole moment in the plasma fireball \cite{Gorbar:2011ya,Burnier:2011bf,Burnier:2012ae} that can explain the experimentally observed \cite{Wang:2012qs,Ke:2012qb} charge-dependent elliptic flows of pions at RHIC \cite{Burnier:2011bf,Burnier:2012ae}. Although
this is quite suggestive to the existence of the phenomenon,
similarly to CME the observable is sensitive to other background effects not originating from triangle anomaly \cite{Dunlop:2011cf,Bzdak:2013yla,Stephanov:2013tga,Deng:2012pc,Huang:2013iia}.

It is desirable to have some observables which are direct consequences of triangle anomaly, yet without having contributions from other backgrounds that have nothing to do with triangle anomaly.
A promising direction is to use discrete symmetries, that is, parity (P) and charge conjugation (C) transformations, to identify such observables, since the axial charge and the triangle anomaly is P- and CP-odd which is a unique characterization of its physics effects. As these discrete symmetries are exact in QCD, any P- and CP-odd observable would be a direct consequence of axial charge fluctuations and the triangle anomaly.

As a first step in this direction, we study possible P- and CP-odd observables in the photon and di-lepton emission rates from a quark-gluon plasma. Since QCD as a theory is P- and CP-even, these
observables are necessarily based on event-by-event P- and CP-odd fluctuations of axial charges.
However, photons and di-leptons are relatively cleaner observables than the hadrons, so our observables may have some potential to be experimentally measurable event-by-event.

Our P- and CP-odd observables for photons and di-leptons from an axially charged plasma
are essentially spin alignments along the momentum direction which measure the net helicity of the photons and di-leptons. Since the axial charge is nothing but the helicity asymmetry of the
fermionic quasi-particles of the plasma \footnote{See Ref.\cite{Loganayagam:2012zg} for an interesting exact relation between fermionic helicity and the axial chemical potential.}, our observables measure how such fermionic helicity is reflected to the helicities of the emitted photons and di-leptons. We will see that our observables are
proportional to the imaginary part of the chiral magnetic conductivity $\sigma_\chi(\omega,k)$
at finite frequency-momentum, defined by a P- and CP-odd part of the retarded current-current correlation functions \cite{Kharzeev:2009pj}
\be
G^{R,-}_{ij}=i\sigma_\chi(\omega,k)\epsilon_{ijk} k^k\,,\quad i,j,k=1,2,3\,,\label{GRodd}
\ee
which is well-known to be a consequence of triangle anomaly. Note that $\sigma_\chi(\omega,k)$
is the coefficient of the CME at finite frequency-momentum,
\be
\vec J=e^2\sigma_\chi(\omega,k)\vec B(\omega,k)\,,
\ee
treating the magnetic field as a linear perturbation with a finite frequency-momentum.
Hence, our observables are the direct tests of the existence of triangle anomaly in QCD.

We emphasize that our observables are not based on the presence of external electromagnetic fields such as the magnetic field in CME/CMW, nor on the geometric flows and anisotropies, and only assume the existence of axial charge fluctuations \footnote{The possible effects of the magnetic field to the photon emission rate, explaining the measured elliptic flow \cite{Adare:2011zr,Lohner:2012ct}, have been discussed in Refs.\cite{Tuchin:2012mf,Basar:2012bp,Fukushima:2012fg,Bzdak:2012fr,Mamo:2012zq,Bu:2013vk,Yee:2013qma}.}. Since the typical relaxation time scale of axial charges in the
RHIC plasma is about $\sim 1-10$ fm depending on  $\alpha_s$, (and larger for LHC with a smaller $\alpha_s$) \cite{Lin:2013sga}, the sign of the net axial charge in a fireball can be coherent event-by-event, and our observables may well be non-zero event-by-event.

An interesting observation on the effect of triangle anomaly to the photons interacting with the plasma was
previously made in Ref.\cite{Akamatsu:2013pjd}, showing that the photon field with a particular polarization is unstable and seems to grow. The physics is based on the same P- and CP-odd part of the retarded correlation functions (\ref{GRodd}), now entering the dispersion relation of photon field interacting with the plasma medium. Although this is quite interesting, for this instability to be realized, the time scale should be long enough to allow multiple interactions between photons and the plasma. Due to a smallness of electromagnetic coupling $\alpha_{\rm EM}\ll1$, this required time scale is parametrically long (proportional to $\alpha_{\rm EM}^{-1}$), and based on this, it has been typically assumed that the photons in heavy-ion collisions once emitted from the plasma do not interact with the plasma again before they leave out the fireball, and the well-known photon emission rate is based on this premise. In this case, the more plausible phenomenon happening in real heavy-ion collisions seems to be a simple asymmetry in the emission rates for different spins we discuss.

\section{P- and CP-odd Observables}

The axial charge in the QCD plasma is a P- and CP-odd quantity, and the experimental signatures from the axially charged plasma should naturally feature some of P- and CP-odd
observables. In this section, we identify such observables in photon and di-lepton emissions, whose experimental measurements may serve as definitive evidences of the existence of axial charges in the plasma created in heavy-ion collisions. Since these observables naturally involve P- and CP-odd part of the charge current correlation functions
which is one of the consequences of the underlying axial-vector-vector triangle anomaly, their observation would also be
a direct evidence of triangle anomaly in QCD.

\subsection{Photons}

Recall that we are considering a homogeneous, isotropic QCD plasma without any external electromagnetic field present. Our only assumption is that the plasma is axially charged, while the vector charge may or may not be present. The axial charge may come from the longitudinal color fields in the early glasma phase, or from thermal sphaleron fluctuations in a later thermalized stage. Since QCD is P- and CP-even (with $\theta_{\rm QCD}=0$), these axial charge fluctuations can only be non-zero event-by-event. Therefore, our P- and CP-odd observables that we discuss in this work should also be taken as event-by-event observables.

We are interested in P- and CP-odd observable in photon emissions from an axially charged isotropic plasma.
Since the plasma is isotropic, let us choose without loss of generality the direction of momentum of the emitted photon to be along $x^3$: $\vec k=k\hat x^3$. What remains is the choice of the polarization of the photon, and it is easy to think of a P-odd quantity associated with photon polarization, which is the circular polarization of the photon.
This is equivalent to the helicity that is whether the
unit spin angular momentum of the photon is along or opposite to the direction of the momentum.
The corresponding polarization vectors for our choice of momentum vector are
\be
\epsilon^\mu_\pm=(\epsilon^0,\epsilon^1,\epsilon^2,\epsilon^3)={1\over \sqrt{2}}(0,1,\pm i,0)\,,\label{polarization}
\ee
where $\pm$ is the helicity of the photon state.
In the presence of axial charge which is P- and CP-odd, the natural and simple observable signal of the axial charge is the difference in the photon emission rates between $+$ and $-$ circularly polarized states, and we define "circular polarization asymmetry",
\be
A_{\pm\gamma}\equiv {{d\Gamma\over d^3 \vec k}\left(\epsilon_+\right)-{d\Gamma\over d^3 \vec k}\left(\epsilon_-\right)\over
{d\Gamma\over d^3 \vec k}\left(\epsilon_+\right)+{d\Gamma\over d^3 \vec k}\left(\epsilon_-\right)}\,,
\ee
where $d\Gamma/d^3 \vec k(\epsilon^\mu)$ is the photon emission rate per unit volume and per unit phase space with a polarization $\epsilon^\mu$.
Since photons are C eigen-states, it is easy to see that $A_{\pm\gamma}$ is P- and CP-odd.
In section \ref{strong}, we provide an exemplar model computation of $A_{\pm\gamma}$ in strongly coupled regime, showing that $A_{\pm\gamma}$ is non-zero if and only if an axial charge is present.

To see how $A_{\pm\gamma}$ probes the P- and CP-odd property of QCD plasma which ultimately comes from the underlying triangle anomaly, let us relate $A_{\pm\gamma}$ with the charge current correlation functions by a well-known formula for the photon emission rate,
\be
{d\Gamma\over d^3 \vec k}\left(\epsilon^\mu\right)={e^2\over (2\pi)^3 2|\vec k|}\epsilon^\mu(\epsilon^\nu)^*G^<_{\mu\nu}(k)\Big|_{k^0=|\vec k|}\,,\label{photonrate}
\ee
with
\be
G^<_{\mu\nu}(k)\equiv \int d^4x\, e^{-ikx}\langle J_\mu(0)J_\nu(x)\rangle\,,\label{gless}
\ee
where $J_\mu$ is the electromagnetic charge current, and our metric convention is $\eta=(-,+,+,+)$.
In the Appendix 1, we present a quantum mechanics derivation of (\ref{photonrate}) as a pedagogic exercise, which
also clarifies how the polarization vector $\epsilon^\mu$ enters the formula.
Relating (\ref{photonrate}) with the retarded correlation functions needs some caution because our polarization vector is complex-valued. Following the steps in the section 2 of Ref.\cite{Yee:2013qma}, we show that the result is
\be
{d\Gamma\over d^3 \vec k}\left(\epsilon^\mu\right)={e^2\over (2\pi)^3 2|\vec k|}{-2\over e^{\beta |\vec k|}-1}{\rm Im}\left[\epsilon^\mu(\epsilon^\nu)^*G^R_{\nu\mu}(k)\right]\Big|_{k^0=|\vec k|}\,,\label{photonrate2}
\ee
with the retarded correlation functions
\be
G^R_{\mu\nu}(k)\equiv -i\int d^4x\,e^{-ikx}\,\theta(x^0)\langle [ J_\mu(x),J_\nu(0)]\rangle\,.
\ee
Note that the polarization vectors are contracted with the retarded correlation functions first before taking the imaginary values.
Using the expression (\ref{polarization}), the polarization-contracted retarded correlation function takes a form
with our choice of the momentum $\vec k=k \hat x^3$ as
\be
\epsilon_\pm^\mu(\epsilon_\pm^\nu)^* G^R_{\nu\mu}={1\over 2}\left(G^R_{11}+G^R_{22}\pm i G^R_{12}\mp iG^R_{21}\right)\,,\label{contractedretarded}
\ee
for the $\pm$ polarized states respectively. Since we have a rotational symmetry in the $(x^1,x^2)$-plane, the correlation functions along $(x^1,x^2)$ must take a form
\be
G^R_{ij}=A \delta_{ij}+B \epsilon_{ij}\,,\quad i,j=1,2\,,
\ee
which dictates that
\be
G^R_{11}=G^R_{22}\,,\quad G^R_{12}=-G^R_{21}\,.\label{rotiden}
\ee
Then, (\ref{contractedretarded}) simplifies to
\be
\epsilon_\pm^\mu(\epsilon_\pm^\nu)^* G^R_{\nu\mu}=\left(G^R_{11}\pm i G^R_{12}\right)\equiv G^R_{\pm}\,,
\ee
in terms of which the circular polarization asymmetry is written as
\be
A_{\pm\gamma}={{\rm Im} \,G^R_+-{\rm Im} \,G^R_-\over {\rm Im} \,G^R_++{\rm Im} \,G^R_-}\Bigg|_{k^0=|\vec k|}={{\rm Re}\, G^R_{12}\over {\rm Im}\, G^R_{11}}\Bigg|_{k^0=|\vec k|}={2{\rm Re}\, G^R_{12}\over {\rm Im}\, {\rm Tr}\, G^R}\Bigg|_{k^0=|\vec k|}=\frac{{\rm Im}\, \sigma_\chi(k^0)}{{\rm Re}\, \sigma_{11}(k^0)}\,,\label{Apmgamma}
\ee
where we used $G^R_{12}\sim ik^0\sigma_\chi(k^0)$ and $G^R_{11}\sim -ik^0\sigma_{11}(k^0)$ to get the last line. We will find the first expression most useful in practical computations later, while the other expressions show that
$A_{\pm\gamma}$ probes a non-vanishing $G^R_{12}$ or a non-vanishing imaginary part of the chiral magnetic conductivity $\sigma_\chi(k^0)$. 

Since in the zero frequency limit, the chiral magnetic conductivity $\sigma_\chi(k^0)$ is given by
\be
\lim_{k^0\rightarrow 0}\sigma_\chi(k^0)={e^2\mu_A\over 2\pi^2}\,,\label{Apmgamma}
\ee
which is real, we expect the imaginary part of the chiral magnetic conductivity hence the circular polarization asymmetry $A_{\pm\gamma}$ to vanish in the zero frequency limit of our numerical computation. We also expect the circular polarization asymmetry $A_{\pm\gamma}$ to be proportional to the axial chemical potential $\mu_A$.

The non-vanishing $G^R_{12}$ when the momentum is $\vec k=k\hat x^3$ is indeed the P- and CP-odd part of the current correlation functions. The only 3D rotationally invariant expression that contributes to $G^R_{12}$ is
\be
G^R_{ij}\sim i\sigma_\chi(k) \epsilon_{ijk} k^k\,,\quad i,j,k=1,2,3\,,\label{GRodd1}
\ee
which is P- and CP-odd. The coefficient $\sigma_\chi(k)$, called the chiral magnetic conductivity \cite{Kharzeev:2009pj}, which is in general a function of $k^\mu$, is responsible for the Chiral Magnetic Effect at finite frequency and momentum, and it is one of the transport phenomena originating from triangle anomaly.
In Refs.\cite{Kharzeev:2009pj,Yee:2009vw}, $\sigma_\chi$ has been computed in weak and strong coupling regimes in the limit $\vec k\to 0$ while $k^0$ is kept finite. Our formula shows that $A_{\pm\gamma}$ measures the imaginary part of the chiral magnetic conductivity
in the kinematic domain of $k^0=|\vec k|$ which hasn't been computed in literature before.

\subsection{Di-leptons}

Let us continue our idea of identifying P- and CP-odd observables in the preceding subsection to the di-lepton emissions from an isotropic axially charged QCD plasma.

We first derive that it is impossible to have P- and CP-odd observables in the di-lepton emission rates if the lepton is strictly massless. Let us recall how P and C transformations act on the leptons and anti-leptons arising from quantizing a single massless Dirac field of the lepton species. For notational familiarity, we will denote leptons as $e^-$ and anti-leptons $e^+$. A single massless Dirac field divides into a left-handed Weyl field $\psi_L$ and a right-handed Weyl field $\psi_R$. Upon quantization, $\psi_L$ produces left-handed leptons $e^-_L$ and right-handed anti-leptons $e^+_R$, while $\psi_R$ field produces right-handed leptons $e^-_R$ and left-handed anti-leptons $e^+_L$.
The P transformation interchanges handedness without affecting the charges (that is $\pm$), and the C transformation interchanges the charges without affecting handedness,
\bear
P &:& e^-_L\leftrightarrow e^-_R\,,\quad e^+_R\leftrightarrow e^+_L\nonumber\\
C &:& e^-_L\leftrightarrow e^+_L\,,\quad e^-_R\leftrightarrow e^+_R\label{PC}
\eear
The crucial point in the argument is that a di-lepton pair is created from a virtual photon whose interaction vertex with
the lepton field does not mix $\psi_L$ and $\psi_R$, that is, the interaction Hamiltonian takes a form
\be
H_I=i e\int d^3 x\, A_\mu \left(\bar\psi_L\gamma^\mu\psi_L+\bar\psi_R\gamma^\mu\psi_R\right)\,.
\ee
This implies that a created di-lepton pair is either $(e^-_L, e^+_R)$ or $(e^-_R, e^+_L)$, and other combinations are forbidden. Now imagine acting P and C transformations on the created di-lepton pair. From (\ref{PC}), we have
\bear
P &:& (e^-_L, e^+_R)\leftrightarrow (e^-_R, e^+_L) \nonumber\\
C &:& (e^-_L, e^+_R)\leftrightarrow (e^+_L, e^-_R)
\eear
which shows that P and C transformations on our restricted set of allowed di-lepton pairs are identical to each other.
It is straightforward to conclude that it is impossible to have something which is P- and CP-odd (which requires to have C-even while P-odd). In our argument, we assumed that the interchange of the momenta of the final lepton and anti-lepton pair has no effect in the emission rate, which is true in an isotropic plasma when the magnitudes of the two momenta are the same and the rate depends only on the relative angle of the two momenta. More sophisticated situations with different magnitudes of momenta of the lepton and anti-lepton might allow some P- and CP-odd observables, but we haven't explored this possibility leaving it as an open question.

The above discussion brings us to consider a massive lepton species. We emphasize that this applies to all known leptons in nature including electrons, but we will see that our proposal for P- and CP-odd observable in di-lepton emission rates is in fact proportional to the mass square of the lepton species, so that it vanishes for a massless species in line with the above discussion. This implies that the signal we propose should be more prominent for heavier lepton species.

\begin{figure}[t]
	\centering
	\includegraphics[width=14cm]{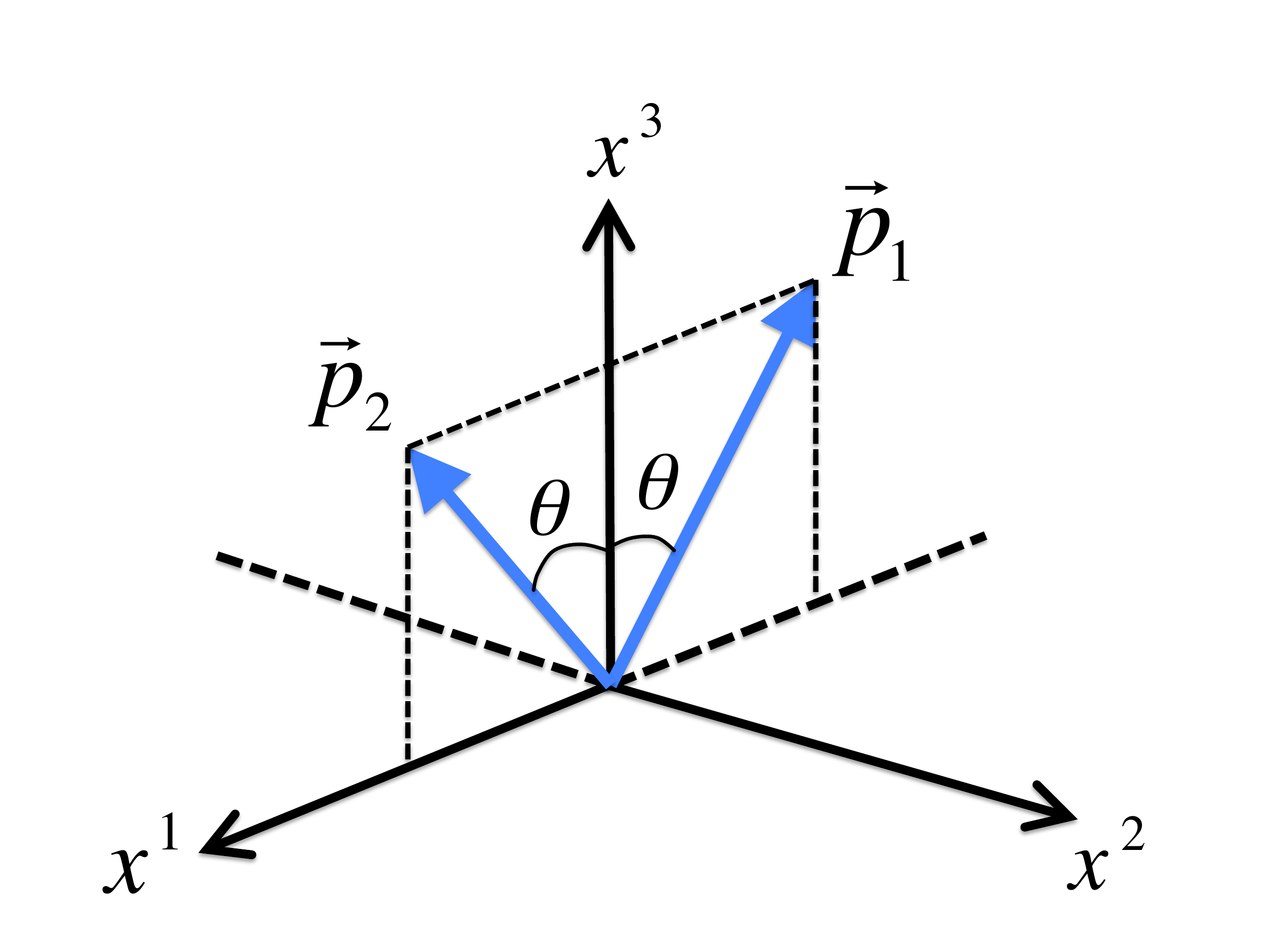}
		\caption{A schematic illustration of the lepton ($\vec p_1$) and anti-lepton ($\vec p_2$) momenta
		in the di-lepton emission from an isotropic axially charged plasma. \label{fig1}}
\end{figure}
Let us define our P- and CP-odd observable in di-lepton emission rates.
We focus on the case where the two momenta of the lepton and anti-lepton have the same magnitude, and form an angle $2\theta$. Calling the two momenta $\vec p_1$ and $\vec p_2$ respectively, we conveniently choose them to be
\be
\vec p_1=p\left(-\sin\theta,0,\cos\theta\right)\,,\quad\vec p_2=p\left(\sin\theta,0,\cos\theta\right)\,,\quad p=|\vec p_1|=|\vec p_2|\,,
\ee
which is schematically depicted in Figure.\ref{fig1}. The total center of mass four-momentum that is carried by the virtual photon is then
\be
p_f^\mu=\left(2E,0,0,2p\cos\theta\right)\,,\quad E=\sqrt{p^2+m^2}\,,
\ee
where $m$ is the mass of the lepton species. The emitted lepton and anti-lepton carry $1/2$ spin degrees of freedom.
For our purpose of discussing parity transformation, it is convenient to choose the helicity basis such that the $+1/2$ state is defined to be a state whose spin along the direction of the momentum is $+1/2$, and similarly for $-1/2$ state.
Each massive lepton and anti-lepton can have the two possible spin states, and
under parity P transformation, the helicity changes its sign.
Let us denote the emission rate per unit volume and per unit phase spaces of lepton and anti-lepton with the spin states given by $s_1$ and $s_2$ respectively by $\Gamma^{s_1,s_2}$,
\be
\Gamma^{s_1,s_2}\equiv {d\Gamma^{s_1,s_2}\over d^3p_1 d^3 p_2}\,,
\ee
and for a given pair of momentum $(\vec p_1,\vec p_2)$, there are four possible rates $\Gamma^{\pm{1\over 2},\pm{1\over 2}}$.
Because of the rotational symmetry of the plasma, the rates with a fixed spin polarization depend only on the relative angle $2\theta$ of the two momenta $(\vec p_1,\vec p_2)$. Under charge conjugation C, the lepton transforms to the anti-lepton without changing helicity, so that the actions of P and C on the spin-polarized emission rates are given by
\bear
P &:& \Gamma^{+{1\over 2},+{1\over 2}}\leftrightarrow  \Gamma^{-{1\over 2},-{1\over 2}}\,,\quad
\Gamma^{+{1\over 2},-{1\over 2}}\leftrightarrow  \Gamma^{-{1\over 2},+{1\over 2}}\nonumber\\
C &:& \Gamma^{\pm{1\over 2},\pm{1\over 2}}\leftrightarrow  \Gamma^{\pm{1\over 2},\pm{1\over 2}}\,,\quad
\Gamma^{+{1\over 2},-{1\over 2}}\leftrightarrow  \Gamma^{-{1\over 2},+{1\over 2}}
\eear
We see that in the subsector of $(\Gamma^{+1/2,-1/2},\Gamma^{-1/2,+1/2})$ (i.e. the opposite spin polarizations for lepton and anti-lepton), the P and C transformations are identical to each other, hence it is impossible to have P and CP-odd observable from that sector. It is however possible to construct a simple P- and CP-odd observable from $\Gamma^{\pm1/2,\pm1/2}$ sector which is
\be
A_{\pm l\bar l}\equiv {\Gamma^{+{1\over 2},+{1\over 2}}-\Gamma^{-{1\over 2},-{1\over 2}}\over
\Gamma^{+{1\over 2},+{1\over 2}}+\Gamma^{-{1\over 2},-{1\over 2}}}\,.
\ee
Note that the total di-lepton rate
\be
\Gamma_{l\bar l}=\Gamma^{+{1\over 2},+{1\over 2}}+\Gamma^{+{1\over 2},-{1\over 2}}+\Gamma^{-{1\over 2},+{1\over 2}}+\Gamma^{-{1\over 2},-{1\over 2}}\,,
\ee
is what used to be computed in literature, and doesn't depend on the choice of the spin basis.

Let us discuss in detail how $A_{\pm l\bar l}$ probes the P- and CP-odd properties of the QCD plasma.
In Appendix 2, we give a mundane quantum mechanics derivation of the di-lepton emission rate with a specified spin polarization $(s_1,s_2)$ given by
\bear
\Gamma^{s_1,s_2}={d\Gamma^{s_1,s_2}\over d^3 p_1 d^3 p_2}&=&{e^2 e_l^2\over (2\pi)^6}
\left(1\over p_f^2\right)^2{1\over 2 E_{\vec p_1}}{1\over 2 E_{\vec p_2}}G^<_{\mu\nu}(p_f)\nonumber\\
&\times&(-1)\left[\bar v(\vec p_2,s_2)\gamma^\mu u(\vec p_1,s_1)\right]\left[\bar u(\vec p_1,s_1)\gamma^\nu v(\vec p_2,s_2)\right]\,,\label{dileptonrate}
\eear
where $E_{\vec p}=\sqrt{\vec p^2+m^2}$, $p_f^\mu=p_1^\mu+p_2^\mu$, the Wightman function $G^<_{\mu\nu}$ is defined in (\ref{gless}) before,
and $u$, $v$ are Dirac spinors of lepton and anti-lepton respectively.
Our convention for the Dirac matrices is
\be
\{\gamma^\mu,\gamma^\nu\}=2\eta^{\mu\nu}\,,\quad \eta=(-1,+1,+1,+1)\,,
\ee
so that $\gamma^0$ is anti-hermitian while $\gamma^i$ ($i=1,2,3$) are hermitian. We will use the following explicit
representation of them
\be
\gamma^0=\left(\begin{array}{cc}0&i{\bf 1}_{2\times 2}\\i{\bf 1}_{2\times 2}& 0\end{array}\right)\,,\quad
\gamma^i=\left(\begin{array}{cc}0&i\sigma^i\\-i\sigma^i& 0\end{array}\right)\,,\quad i=1,2,3\,,
\ee
upon which the spin matrix corresponding to a spatial rotation of angle $\theta$ along an axis $\hat n$ is given simply by
\be
S(\hat n,\theta)=e^{i\theta \vec S\cdot\hat n}\,,\quad S^i\equiv{1\over 2}\left(\begin{array}{cc} \sigma^i &0\\0&\sigma^i\end{array}\right)\,.
\ee
Recall that the above acts on the lepton wave functions, both positive and negative energy states. For the anti-leptons which are holes of the lepton wave functions with negative energy, the actual spin matrix is in fact a negative of the above.
Note also that our definition for $\bar\psi$ is $\bar\psi\equiv -i\psi^\dagger \gamma^0$, and the Dirac equation is
$(\gamma^\mu\partial_\mu-m)\psi=0$.
With the above conventions, the Dirac spinors are given by
\be
u(\vec p,s)=\left(\begin{array}{c} \sqrt{E-\vec p\cdot\vec\sigma}\xi^s\\\sqrt{E+\vec p\cdot\vec\sigma}\xi^s\end{array}\right)\,,\quad
v(\vec p,s)=\left(\begin{array}{c} \sqrt{E-\vec p\cdot\vec\sigma}\eta^s\\-\sqrt{E+\vec p\cdot\vec\sigma}\eta^s\end{array}\right)\,,\label{diracspinors}
\ee
where
\be
(\vec p\cdot\vec\sigma) \xi^s=(2s)|\vec p|\xi^s\,,\quad (\vec p\cdot\vec\sigma) \eta^s=-(2s)|\vec p|\eta^s\,,\quad s=\pm{1\over 2}\,.
\ee
With our choice of the momenta $\vec p_1$ and $\vec p_2$ as in Figure.\ref{fig1}, the Dirac spinors are explicitly
found to be
\be
u\left(\vec p_1,+1/2 \right)=\left(\begin{array}{c}+\sqrt{E-p}\cos(\theta/2)\\-\sqrt{E-p}\sin(\theta/2)\\
+\sqrt{E+p}\cos(\theta/2)\\-\sqrt{E+p}\sin(\theta/2)\end{array}\right)\,,\quad
u\left(\vec p_1,-1/2 \right)=\left(\begin{array}{c}+\sqrt{E+p}\sin(\theta/2)\\+\sqrt{E+p}\cos(\theta/2)\\
+\sqrt{E-p}\sin(\theta/2)\\+\sqrt{E-p}\cos(\theta/2)\end{array}\right)\,,
\ee
and
\be
v\left(\vec p_2,+1/2 \right)=\left(\begin{array}{c}-\sqrt{E+p}\sin(\theta/2)\\+\sqrt{E+p}\cos(\theta/2)\\
+\sqrt{E-p}\sin(\theta/2)\\-\sqrt{E-p}\cos(\theta/2)\end{array}\right)\,,\quad
v\left(\vec p_2,-1/2 \right)=\left(\begin{array}{c}+\sqrt{E-p}\cos(\theta/2)\\+\sqrt{E-p}\sin(\theta/2)\\
-\sqrt{E+p}\cos(\theta/2)\\-\sqrt{E+p}\sin(\theta/2)\end{array}\right)\,.
\ee

With the above expressions, let us then compute the spin-polarization contracted Wightman function that the emission rate (\ref{dileptonrate}) is proportional to
\bear
G^{s_1,s_2}&\equiv&
(-1)G^<_{\mu\nu}(p_f)\left[\bar v(\vec p_2,s_2)\gamma^\mu u(\vec p_1,s_1)\right]\left[\bar u(\vec p_1,s_1)\gamma^\nu v(\vec p_2,s_2)\right]\nonumber\\
&=&{2\over e^{\beta p_f^0}-1}{\rm Im}\left[G^R_{\mu\nu}(p_f)\left[\bar v(\vec p_2,s_2)\gamma^\mu u(\vec p_1,s_1)\right]\left[\bar u(\vec p_1,s_1)\gamma^\nu v(\vec p_2,s_2)\right]\right]\,,
\eear
where the second line is obtained from the standard manipulation with Lehmann representation, using the fact that what multiplies to $G^<_{\mu\nu}$ from Dirac spinors is hermitian with respect to $\mu\nu$ indices (see the section 2 in Ref.\cite{Yee:2013qma}).
After some algebra, we find
\bear
G^{\pm{1\over 2},\pm{1\over 2}}&=&4m^2\left((1+\cos^2\theta)G^<_{11}\pm2i\cos\theta G^<_{12}\right)\nonumber\\
&=&2m^2\left((1\pm\cos\theta)^2\left(G^<_{11}+iG^<_{12}\right)+(1\mp\cos\theta)^2\left(G^<_{11}-iG^<_{12}\right)\right)\nonumber\\
&=&{-4m^2\over e^{\beta p_f^0}-1}
\left((1\pm\cos\theta)^2{\rm Im}\left(G^R_{11}+iG^R_{12}\right)+(1\mp\cos\theta)^2{\rm Im}\left(G^R_{11}-iG^R_{12}\right)\right)\nonumber\\
&=&{-4m^2\over e^{\beta p_f^0}-1}
\left((1\pm\cos\theta)^2{\rm Im}\,G^R_++(1\mp\cos\theta)^2{\rm Im}\,G^R_-\right)\,,
\eear
where we have used $G^R_{11}=G^R_{22}$ and $G^R_{12}=-G^R_{21}$ from the rotational symmetry as in (\ref{rotiden}), and $G^R_\pm\equiv(G^R_{11}\pm i G^R_{12})$ as defined before.
It is possible to compute $G^{\pm1/2,\mp1/2}$ too, but the results are not of our interests.
They are given by
\be
G^{+{1\over 2},-{1\over 2}}=G^{-{1\over 2},+{1\over 2}}=4\left(p^2\cos^2\theta-E^2\right)\left(G^<_{00}-G^<_{33}\right)+4E^2\sin^2\theta G^<_{11}\,,
\ee
after using the Ward identity
\be
E G^<_{00}+p\cos\theta G^<_{30}=0\,,\quad EG^<_{03}+p\cos\theta G^<_{33}=0\,.
\ee
We confirm that $G^{\pm 1/2,\pm 1/2}$ are proportional to the mass square of the lepton species, so that our P- and CP-odd observable $A_{\pm l\bar l}$ constructed from them is well-defined for a massive species only.

Using the above results, our final expression for P- and CP-odd observable $A_{\pm l\bar l}$ in the di-lepton emission rate in terms of retarded correlation functions is
\be
A_{\pm l\bar l}=\left({2\cos\theta\over 1+\cos^2\theta}\right)\cdot{{\rm Im} \,G^R_+-{\rm Im} \,G^R_-\over {\rm Im} \,G^R_++{\rm Im} \,G^R_-}\Bigg|_{p^\mu=p^\mu_f=p_1^\mu+p_2^\mu}\,,\label{Apmdilepton}
\ee
which is similar to the expression (\ref{Apmgamma}) for $A_{\pm\gamma}$, except an additional angular factor and a different
kinematic domain probed. We see that $A_{\pm l\bar l}$ is also a consequence of the underlying triangle anomaly, and it measures the chiral magnetic conductivity in a different kinematic domain.

In the next section, we give an exemplar computation of ${\rm Im}\,G^R_\pm$ in strongly coupled regime, using AdS/CFT correspondence, and present some numerical results for our P- and CP-odd observables $A_{\pm\gamma}$ and $A_{\pm l\bar l}$ that may be relevant in realistic heavy-ion experiments.

\section{Strong Coupling Computation\label{strong}}

The purpose of this section is to present one exemplar model computation of our P- and CP-odd
observables $A_{\pm\gamma}$ and $A_{\pm l\bar l}$ in strongly coupled regime using AdS/CFT correspondence. Since the chiral symmetry, especially the triangle anomaly represented by 5 dimensional Chern-Simons terms, is an important ingredient in our observables, the holographic model to be used should describe the right chiral symmetry in the real QCD.
Instead of using a bottom-up approach, we choose to work in the Sakai-Sugimoto model~\cite{Sakai:2004cn} which is the only top-down holographic model whose chiral symmetry is identical to the one in QCD \footnote{
We should bear in mind that the background holographic space-time from the D4 branes in the deconfined phase
of the model has some problem with center symmetry, which prevents us from relating it to the true gluonic sector of the real QCD \cite{Mandal:2011ws}. However, this problem is absent for the chiral symmetry dynamics described by the probe D8 branes that our analysis is focused on.}.

For our purposes, it is enough to start from the following description of the model in its finite temperature deconfined phase. Our presentation is oriented only for its practical usage skipping details of its derivations (For a more complete description, see for example the section 5 of Ref.\cite{Yee:2009vw} and section 3 of Ref.\cite{Yee:2013qma}.) We consider the case of having a single massless Dirac quark species whose electric charge is $e$.
The model lives in a 5 dimensional space-time, $(x^\mu,U)$ where $U$ is an extra holographic dimension. There are two 5 dimensional U(1) gauge fields, $A_V$ and $A_a$, corresponding to
the vector and axial symmetry of the massless quark species in the QCD side, whose 5 dimensional dynamics describes the chiral dynamics of the massless quark holographically.
Especially, there are 5 dimensional Chern-Simons terms that are the holographic manifestation of the triangle anomaly in the QCD side
\be
S_{\rm CS}={N_c\over 96\pi^2}\int d^4x dU\,\epsilon^{MNPQR}\left[-(A_L)_M(F_L)_{NP}(F_L)_{QR}
+(A_R)_M(F_R)_{NP}(F_R)_{QR}\right]\,,
\ee
where we introduce chiral gauge fields defined by
\be
A_L=A_V-A_a\,,\quad A_R=A_V+A_a\,.
\ee
The QCD plasma with a finite axial charge is described in the model by a non-zero background configuration of the axial gauge field $A_a$ which is \footnote{We put $2\pi l_s^2=1$ in the formulas from Ref.\cite{Yee:2009vw} for convenience, since its value doesn't appear in the final QCD observables.}
\be
(F_a)_{tU}^{(0)}=-{\alpha\over\sqrt{U^5+\alpha^2}}\,,\label{Fa}
\ee
where the parameter $\alpha$ is related to the axial chemical potential $\mu_A$ by the relation
\be
\mu_A=\int_{U_T}^\infty dU\, {\alpha\over\sqrt{U^5+\alpha^2}}={2\alpha\over 3 U_T^{3\over 2}}    {_2F_1}\left({3\over 10},{1\over 2},{13\over 10},-{\alpha^2\over U_T^{5}}\right)\,.
\ee
The parameter $U_T$ in the above in turn is determined by the temperature $T$ by
\be
U_T=R^3\left(4\pi T\over 3\right)^2\,,
\ee with a numerical value $R^3=1.44$ in units of GeV.
The $U_T$ is in fact the location of the black-hole horizon at $U=U_T$ in the background holographic space-time describing a finite temperature plasma, and the holographic coordinate $U$ has a range
$U_T < U<\infty$ where $U=\infty$ is the region corresponding to the UV regime of the QCD side.

Our main interest is to compute retarded (vector) current correlation functions in the axially charged
plasma described above. To do this in holography, one first solves the linearized equations of
motion for the vector gauge field $A_V$ fluctuations from the background solution given by (\ref{Fa}) \cite{Yee:2009vw}
\bear
&&\partial_U(A(U)F_{tU})-B(U)(\partial_i F_{Ui})=0\,,\label{lineareom}\\
&&A(U)(\partial_t F_{tU})+B(U)(\partial_i F_{ti})+C(U)(\partial_i F_{Ui})=0\,,\nonumber\\
&&B(U)(\partial_t F_{Ui})+\partial_U(B(U) F_{ti}+C(U) F_{Ui})+D(U)\partial_j F_{ji}-{N_c\over 8\pi^2 C}(F_a)^{(0)}_{tU} \epsilon^{ijk} F_{jk}=0\,,\nonumber
\eear
where $i,j,k=1,2,3$, $C=0.0211$ in units of GeV, and the functions $A(U), B(U), C(U), D(U)$ are given by
\bear
A(U)&=&U^{-5}\left(U^5+\alpha^2\right)^{3\over 2}\,,\quad
B(U)= \left(R\over U\right)^{3\over 2}\left(U^5+\alpha^2\right)^{1\over 2}\,,\nonumber\\
C(U)&=&f(U)\left(U^5+\alpha^2\right)^{1\over 2}\,,\quad
D(U)= \left(R\over U\right)^3 U^5\left(U^5+\alpha^2\right)^{-{1\over 2}}\,,
\eear
with
\be
f(U)=1-\left(U_T\over U\right)^3\,.
\ee
Note that the last term in the third equation in (\ref{lineareom}) is from the 5 dimensional Chern-Simons term which is a consequence of triangle anomaly. The solution has a near $U\to\infty$ behavior given by
\be
A_\mu=A^{(0)}_\mu +{A^{(1)}_\mu\over U^{1\over 2}}+{A^{(2)}_\mu\over U}+{\tilde A_\mu\over U^{{3\over 2}}}+\cdots\,,\label{UV}
\ee
with
\bear
A^{(1)}_t=0\,,\quad A^{(1)}_i=2R^{3\over 2}F_{ti}^{(0)}\,,\quad A^{(2)}_t=-2R^3 \partial_j F^{(0)}_{tj}\,,\quad
A^{(2)}_i=-2R^3 \partial_j F^{(0)}_{ij}\,,
\eear
where $A^{(0)}_\mu$ is a free parameter (the UV boundary condition) acting as a source for the QCD vector current $J^\mu$, while the $\tilde A_\mu$ is a dynamically determined quantity which encodes
the expectation value of the current in the presence of the source $A_\mu^{(0)}$ by \cite{Yee:2013qma}
\bear
\langle J_t\rangle&=& 3C\left(\tilde A_t+{8\over 3}R^{9\over 2}\partial_t\partial_j F^{(0)}_{tj}\right)\,,\nonumber\\
\langle J_i\rangle&=& 3C\left(\tilde A_i+4R^{9\over 2}\left(\partial_t\partial_j F^{(0)}_{ij}+{2\over 3}\partial_t^2 F^{(0)}_{ti}-{1\over 3}\partial_i\partial_j F^{(0)}_{tj}\right)\right)\,.\label{expectation}
\eear
The solution with a given source $A_\mu^{(0)}$ and the incoming boundary condition at the horizon $U=U_T$ is unique and it is proportional to $A_\mu^{(0)}$, and hence the current expectation value
(\ref{expectation}) is a linear function of $A_\mu^{(0)}$ from which we finally obtain our desired retarded correlation functions as
\be
\langle J_\mu\rangle = -G^{R\,\,\nu}_\mu A^{(0)}_\nu\,.
\ee

Since we are interested in computing only the transverse part of the correlation functions, we can
consistently turn on $A_{1,2}$ components only, after taking the frequency-momentum $(\omega,\vec k=k\hat x^3)$, so that $\partial_t=-i\omega$, $\partial_i=ik\delta_{i3}$. The relevant equation of motion is the third equation in (\ref{lineareom}),
\be
-i\omega B(U)\partial_U A_i+\partial_U\left(-i\omega B(U) A_i+C(U)\partial_U A_i\right)
-k^2 D(U)A_i +ik{N_c\over 8\pi^2 C}(F_a)^{(0)}_{tU}\epsilon^{ij}A_j=0\,,
\ee
with $i,j=1,2$ and $\epsilon^{12}=-\epsilon^{21}=+1$.
From the structure of the above equation, it is natural to work with a helicity basis
\be
A_\pm={1\over\sqrt{2}}\left(A_1\mp i A_2\right)\,,
\ee
in terms of which the equation of motion diagonalizes as
\be
-i\omega B(U)\partial_U A_\pm+\partial_U\left(-i\omega B(U) A_\pm+C(U)\partial_U A_\pm\right)
-k^2 D(U)A_\pm \mp k{N_c\over 8\pi^2 C}(F_a)^{(0)}_{tU}A_\pm=0\,.\label{diagonal}
\ee
Once we find the solution of $A_\pm$, we can read off the source $A^{(0)}_\pm=1/\sqrt{2}(A^{(0)}_1\mp A^{(0)}_2)$ and the expectation
value via (\ref{expectation})
\be
\langle J^\pm\rangle ={1\over\sqrt{2}}\left(J^1\mp i J^2\right)=3C\left(\tilde A_\pm+4R^{9\over 2}(-i\omega)\left(k^2-{2\over 3}\omega^2\right) A^{(0)}_\pm\right)\,.\label{exp}
\ee
From the relation $\langle J^i\rangle=-G^{Rij}A^{(0)}_j$, and the rotational symmetry $G^R_{11}=G^R_{22}$ and $G^R_{12}=-G^R_{21}$, it is straight forward to see that
\be
\langle J^\pm\rangle = - \left(G^R_{11}\pm i G^R_{12}\right) A^{(0)}_\pm=-G^R_\pm A^{(0)}_\pm\,,
\ee
so that we can naturally obtain our desired $G^R_\pm$, entering our expressions (\ref{Apmgamma}) and (\ref{Apmdilepton}) for $A_{\pm\gamma}$ and $A_{\pm l\bar l}$, from the solutions of $A_\pm$.

Numerically, what we do is to solve the equation (\ref{diagonal}) from the horizon $U=U_T$
up to a UV maximum $U_{max}$ and then compare its value and derivative at $U_{max}$ with the UV expansion (\ref{UV}),
\bear
A_\pm(U_{max})&=&A^{(0)}_\pm + { 2 R^{3\over 2}(-i\omega)\over U_{\max}^{1\over 2}}A_\pm^{(0)}+{-2 R^3 k^2\over U_{\max}}A^{(0)}_\pm +{\tilde A_\pm\over U_{\max}^{3\over 2}}\,,\nonumber\\
\partial_U A_\pm(U_{\max})&=& -{1\over 2}{ 2 R^{3\over 2}(-i\omega)\over U_{\max}^{3\over 2}}A_\pm^{(0)}+{2 R^3 k^2\over U_{\max}^2}A^{(0)}_\pm-{3\over 2}{\tilde A_\pm\over U_{\max}^{5\over 2}}\,,
\eear
to obtain $A_\pm^{(0)}$ and $\tilde A_\pm$. We then compute $\langle J^\pm\rangle$ from (\ref{exp}),
and finally get $G^R_\pm$ from
\be
G^R_\pm=-{\langle J^\pm\rangle\over A^{(0)}_\pm}\,.
\ee

\begin{figure}[t]
	\centering
	\includegraphics[width=14cm]{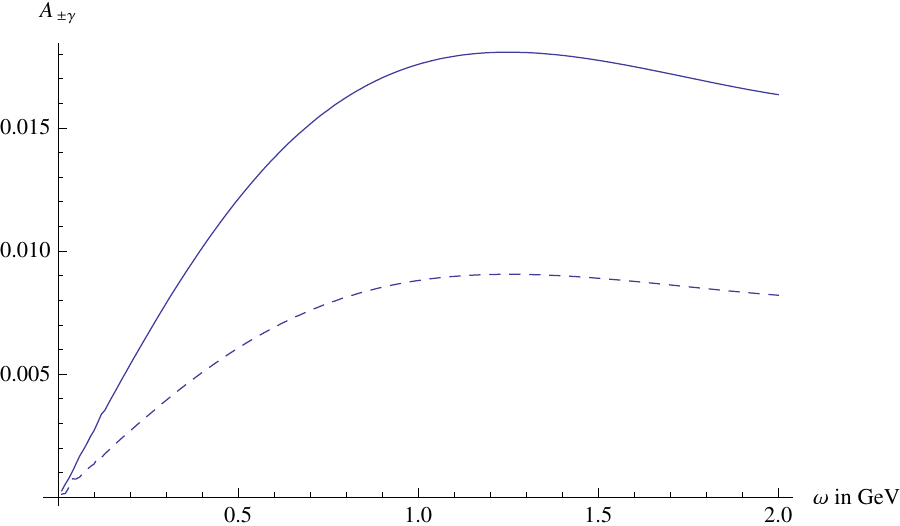}
		\caption{The photon circular polarization asymmetry $A_{\pm\gamma}$ from an axially charged plasma as a function of frequency $\omega$, where $T=300$ MeV with $\mu_A=100$ MeV (solid) and $\mu_A=50$ MeV (dashed).\label{fig2}}
\end{figure}
\begin{figure}[t]
	\centering
	\includegraphics[width=14cm]{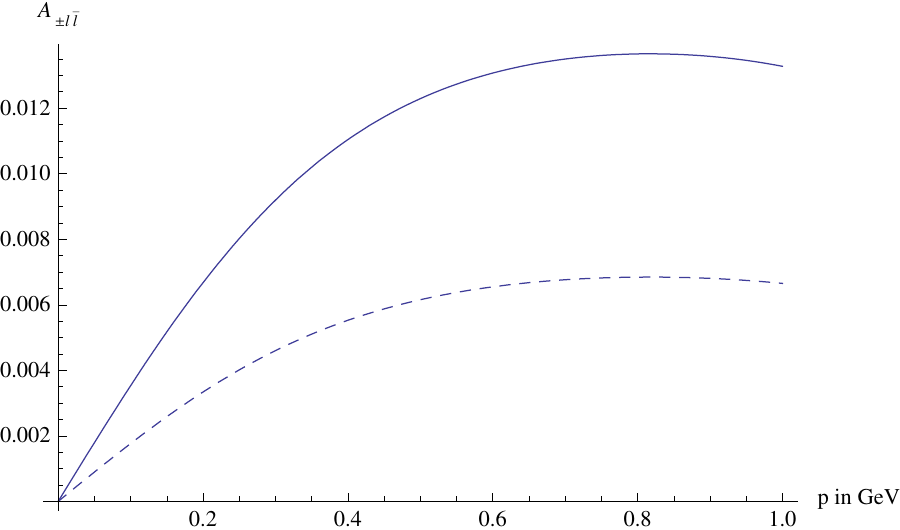}
		\caption{The di-lepton spin polarization asymmetry $A_{\pm l\bar l}$ from an axially charged plasma as a function of one lepton momentum $p=|\vec p|$ for the case of muon, where $T=300$ MeV with $\mu_A=100$ MeV (solid) and $\mu_A=50$ MeV (dashed). The relative angle between muon and anti-muon pair is taken to be $2\theta={\pi\over 2}$. \label{fig3}}
\end{figure}
Figure \ref{fig2} shows our numerical results of photon circular polarization asymmetry $A_{\pm\gamma}$ as a function of frequency, where $T=300$ MeV with $\mu_A=100$ MeV (solid) and $\mu_A=50$ MeV (dashed).
Since the model is trustable only up to a few GeV's, we compute $A_{\pm\gamma}$ only for $\omega < 2$ GeV. We observe that the asymmetry is about a percent level with a peak around $\omega=1$ GeV.
It is easy to check that the result is absent without the Chern-Simons term (triangle anomaly) and the effect is roughly proportional to the axial chemical potential.

Figure \ref{fig3} shows our numerical results for the di-lepton spin polarization asymmetry $A_{\pm l\bar l}$
in the case of di-muon pair with a relative angle $2\theta={\pi\over 2}$ as a function of the muon momentum $p=|\vec p|$ (see Figure \ref{fig1}). Note that the $p_f^\mu$  which probes the plasma is
\be
p_f^0=2\sqrt{p^2+m_\mu^2}\,,\quad m_\mu=100\,{\rm MeV}\,,\quad |\vec p_f|=2p\cos\theta \,.
\ee
We observe again that the effect is about a percent level.

\section{Discussion}

In this work, we identify P- and CP-odd observables in the photons and di-lepton emission rates from an axially charged isotropic QCD plasma, whose experimental observation can be a direct confirmation
of the triangle anomaly in QCD.
We show that these observables are proportional to the imaginary part of the chiral magnetic conductivity at a finite frequency-momentum region, which is originating from the underlying triangle anomaly.
With an ideal set of parameters of the temperature and the axial chemical potential, our exemplar model computation in AdS/CFT correspondence shows that our observables are of a percent level.
Clearly, a more realistic estimate of the net effect is desirable, including various real time fireball dynamics and the axial charge fluctuation dynamics in both early glasma phase and the later quark-gluon plasma stage.
Another direction is to compute our observables in different models such as weak coupling chiral kinetic theory that has been recently developed in Refs.\cite{Son:2012wh,Stephanov:2012ki,Chen:2012ca}.

\vskip 1cm \centerline{\large \bf Acknowledgment} \vskip 0.5cm

We thank Adam Bzdak, Dima Kharzeev, Shu Lin, Larry McLerran, Vlad Skokov, Misha Stephanov, and Raju Venugopalan for interesting discussions.

\section*{Appendix 1 : Photon Emission Formula from Quantum Mechanics}

We derive the photon emission rate formula (\ref{photonrate}) in terms of the Wightman function $G^<_{\mu\nu}$,
\be
{d\Gamma\over d^3 \vec k}\left(\epsilon^\mu\right)={e^2\over (2\pi)^3 2|\vec k|}\epsilon^\mu(\epsilon^\nu)^*G^<_{\mu\nu}(k)\Big|_{k^0=|\vec k|}\,,\label{photonappendix}
\ee
in the framework of quantum mechanics perturbation theory.
The Hilbert space of our interest is a tensor product of the QCD plasma Hilbert space and the Hilbert space of photons,
\be
{\cal H}={\cal H}_{\rm QCD}\otimes{\cal H}_\gamma\,.
\ee
The photon emission process from a QCD plasma is a quantum mechanical transition from an initial state
\be
|i\rangle=|\alpha_i\rangle\otimes|0\rangle\,,\label{initial}
\ee
to a final state containing one photon quantum with a momentum $\vec k$ and a polarization $\epsilon^\mu$,
\be
|f\rangle=|\alpha_f\rangle\otimes|\vec k,\epsilon^\mu\rangle\,,\label{final}
\ee
where $|\alpha_{i,f}\rangle$ are QCD states describing the plasma. We will perform a thermal ensemble average for the initial QCD state $|\alpha_i\rangle$, while summing over all possible final QCD states $|\alpha_f\rangle$ to get a final emission rate for a fixed photon state $|\vec k,\epsilon^\mu\rangle$.

The transition from our initial state to the final state with one photon arises due to an interaction Hamiltonian
\be
H_I=e\int d^3x\, A_\mu(\vec x,t=0)J^\mu(\vec x,t=0)\,,
\ee
where $A_\mu$ is the photon field operator and $J^\mu$ is the electromagnetic current operator in the QCD sector.
Note that $A_\mu$ acts only on ${\cal H}_\gamma$, and $J^\mu$ on ${\cal H}_{\rm QCD}$ only.
Since it is easy to see that the matrix element of $H_I$ between our initial and final states is non-zero,
the transition process is described by a first order perturbation theory where the transition rate is given by the
Fermi's Golden rule
\be
T_{i\to f}=(2\pi)\left|\langle f|H_I|i\rangle\right|^2 \delta(E_f-E_i)\,,\label{Fermi1}
\ee
where $E_{i,f}$ are the energies of the initial and final states. In this expression, it is important to have the right normalization for the states, $\langle i|i\rangle=\langle f|f\rangle=1$.
To keep track of normalization of the states properly, we work in a finite volume case
throughout our derivation until we take an infinite volume limit at the end.
A spatial momentum in a finite volume $V$ is discrete and labeled by a triple of integers $\vec n$, so that
we denote it as $\vec k_{\vec n}$.
The number of such momentum states in a phase space volume $d^3 k$ is well-known to be $V{d^3 k\over(2\pi)^3}$.

The photon field operator $A_\mu(\vec x,t)$ in a finite volume $V$ has a standard expansion in terms of
creation and annihilation operators of individual photon states with momenta $\vec k_{\vec n}$ and polarization $\epsilon_\mu^{(s)}$ as
\be
A_\mu(\vec x,t)={1\over V}\sum_{\vec n,s}{1\over \sqrt{2|\vec k_{\vec n}|}}\left(\epsilon_\mu^{(s)} a^{(s)}_{\vec k_{\vec n}}\e^{-i|\vec k_{\vec n}|t+i\vec k_{\vec n}\cdot\vec x}+{\rm h.c}\right)\,,\label{photonquantum}
\ee
where the creation/annihilation operators satisfy
\be
[a^{(s)}_{\vec k_{\vec n}}, a^{(s')\dagger}_{\vec k_{\vec n'}}]=V\delta_{\vec n, \vec n'}\delta^{s,s'}\,.
\ee
In the above, we are careful about how the volume factors enter to have a right normalization of the field commutation relations. With this, the properly normalized one photon state in the Hilbert space $\cal{H}_\gamma$ is
\be
|\vec k_{\vec n},s\rangle={1\over\sqrt{V}}a^{(s)\dagger}_{\vec k_{\vec n}}|0\rangle\,.
\ee

It is then straightforward to compute the matrix element $\langle f|H_I|i\rangle$ for our initial and final states (\ref{initial}) and (\ref{final}). Using
\be
\langle \vec k_{\vec n},s|A_\mu(\vec x,t=0)|0\rangle={1\over\sqrt{V}}{1\over\sqrt{2|\vec k_{\vec n}|}}\epsilon_\mu^{(s)*}e^{-i\vec k_{\vec n}\cdot\vec x}\,,
\ee
we have
\be
\langle f|H_I|i\rangle={1\over\sqrt{V}}{e\over\sqrt{2|\vec k_{\vec n}|}}\epsilon_\mu^{(s)*}\int d^3 x\,\langle\alpha_f| J^\mu(\vec x,t=0)|\alpha_i\rangle e^{-i\vec k_{\vec n}\cdot\vec x}\,.
\ee
Denoting the energies of the QCD states $|\alpha_{i,f}\rangle$ by $\varepsilon_{i,f}$, the initial and final state energies are $E_i=\varepsilon_i$ and $E_f=\varepsilon_f+|\vec k_{\vec n}|$, so that the transition rate by (\ref{Fermi1}) becomes
\bear
T_{i\to f}&=&{(2\pi)\over V}{e^2\over 2|\vec k_{\vec n}|}\epsilon_\mu^{(s)*}\epsilon_\nu^{(s)}\delta(\varepsilon_f-\varepsilon_i+|\vec k_{\vec n}|)\nonumber\\&\times&\int d^3 x\int d^3 y \,e^{i\vec k_{\vec n}\cdot(\vec y-\vec x)}\langle \alpha_i|J^\nu(\vec y,t=0)|\alpha_f\rangle\langle\alpha_f| J^\mu(\vec x,t=0)|\alpha_i\rangle\,.
\eear
We now do the following manipulation on the above. We first replace $\delta(\varepsilon_f-\varepsilon_i+|\vec k_{\vec n}|)$ by
\be
\delta(\varepsilon_f-\varepsilon_i+|\vec k_{\vec n}|)={1\over 2\pi}\int dt\, e^{it(\varepsilon_f-\varepsilon_i+|\vec k_{\vec n}|)}\,,
\ee
and combine the factor $e^{it(\varepsilon_f-\varepsilon_i)}$ from the above with $\langle\alpha_f| J^\mu(\vec x,t=0)|\alpha_i\rangle$
to have
\be
e^{it(\varepsilon_f-\varepsilon_i)}\langle\alpha_f| J^\mu(\vec x,t=0)|\alpha_i\rangle=\langle\alpha_f| J^\mu(\vec x,t)|\alpha_i\rangle\,,
\ee
using the fact that $J^\mu(\vec x,t)=e^{iHt}J^\mu(\vec x,t=0) e^{-iHt}$. The result is
\bear
T_{i\to f}&=&{1\over V}{e^2\over 2|\vec k_{\vec n}|}\epsilon_\mu^{(s)*}\epsilon_\nu^{(s)}\nonumber\\&\times&\int d^3 x\int d^3 y \int dt \,e^{i|\vec k_{\vec n}|t+i\vec k_{\vec n}\cdot(\vec y-\vec x)}\langle \alpha_i|J^\nu(\vec y,t=0)|\alpha_f\rangle\langle\alpha_f| J^\mu(\vec x,t)|\alpha_i\rangle\,.
\eear
We then sum over the final QCD states $|\alpha_f\rangle$ to remove $\sum_f |\alpha_f\rangle\langle\alpha_f|=1$ in the middle, and perform a thermal ensemble average over $\alpha_i$,
\be
\langle J^\nu(\vec y,t=0) J^\mu(\vec x,t)\rangle\equiv {1\over Z}\sum_i e^{-\beta\varepsilon_i}
\langle \alpha_i|J^\nu(\vec y,t=0)J^\mu(\vec x,t)|\alpha_i\rangle\,,
\ee
to have
\be
T_{i\to f}={1\over V}{e^2\over 2|\vec k_{\vec n}|}\epsilon_\mu^{(s)*}\epsilon_\nu^{(s)}
\int d^3 x\int d^3 y \int dt \,e^{i|\vec k_{\vec n}|t+i\vec k_{\vec n}\cdot(\vec y-\vec x)}\langle J^\nu(\vec y,t=0)J^\mu(\vec x,t)\rangle\,.
\ee
Exploring the translational symmetry of the plasma such that $\langle J^\nu(\vec y,t=0)J^\mu(\vec x,t)\rangle$ depends only on the relative displacement $\vec x-\vec y$, one can simply replace $\vec x-\vec y$ with $\vec x$ in the integrand while getting additional volume factor $\int d^3 y=V$, to have
\bear
T_{i\to f}&=&{e^2\over 2|\vec k_{\vec n}|}\epsilon_\mu^{(s)*}\epsilon_\nu^{(s)}
\int d^3 x\int dt \,e^{i|\vec k_{\vec n}|t-i\vec k_{\vec n}\cdot\vec x}\langle J^\nu(\vec 0,t=0)J^\mu(\vec x,t)\rangle\nonumber\\&=&{e^2\over 2|\vec k_{\vec n}|}\epsilon^{(s)\mu*}\epsilon^{(s)\nu}G^<_{\nu\mu}(k^0=|\vec k_{\vec n}|,\vec k_{\vec n})\,,
\eear
where the Wightman function is defined as before
\be
G^<_{\mu\nu}(k)=\int d^4x\, e^{-ikx}\langle J_\mu(0)J_\nu(x)\rangle\,.
\ee
Recalling that the number of momentum states within a phase space volume $d^3 k$ is $V{d^3 k\over (2\pi)^3}$,
the total transition rate to the states having one photon within a phase space volume $d^3 k$ is obtained by multiplying the above $T_{i\to f}$ by $V{d^3 k\over (2\pi)^3}$. Then,
the photon emission rate per unit volume and per unit phase space volume is
\be
{d\Gamma\over d^3 k}={1\over (2\pi)^3}T_{i\to f}=
{e^2\over (2\pi)^3 2|\vec k_{\vec n}|}\epsilon^{(s)\mu*}\epsilon^{(s)\nu}G^<_{\nu\mu}(k^0=|\vec k_{\vec n}|,\vec k_{\vec n})\,,
\ee
which is our desired formula (\ref{photonappendix}) after taking an infinite volume limit to replace discrete $\vec k_{\vec n}$ with a continuum $\vec k$.

\section*{Appendix 2 : Di-lepton Emission Formula from Quantum Mechanics}

We would like to give a quantum mechanics derivation of the di-lepton emission formula (\ref{dileptonrate}),
\bear
{d\Gamma^{s_1,s_2}\over d^3 p_1 d^3 p_2}&=&{e^2 e_l^2\over (2\pi)^6}
\left(1\over p_f^2\right)^2{1\over 2 E_{\vec p_1}}{1\over 2 E_{\vec p_2}}G^<_{\mu\nu}(p_f)\nonumber\\
&\times&(-1)\left[\bar v(\vec p_2,s_2)\gamma^\mu u(\vec p_1,s_1)\right]\left[\bar u(\vec p_1,s_1)\gamma^\nu v(\vec p_2,s_2)\right]\,,\label{dileptonappendix}
\eear
where $p_f=p_1+p_2$ is the total di-lepton energy-momentum and $e_l$ is the electric charge of the lepton species.
The Hilbert space of our interests consists of three parts, the QCD sector ${\cal H}_{\rm QCD}$, the photon sector ${\cal H}_\gamma$, and the lepton sector ${\cal H}_l$ : ${\cal H}={\cal H}_{\rm QCD}\otimes{\cal H}_\gamma\otimes{\cal H}_l$.
The di-lepton emission process is a transition from the initial state
\be
|i\rangle=|\alpha_i\rangle\otimes|0\rangle\otimes|0\rangle\,,\label{initial2}
\ee
to a final state containing the lepton and anti-lepton pair with momenta $\vec p_{1,2}$ and spin polarizations $s_{1,2}$ respectively,
\be
|f\rangle=|\alpha_f\rangle\otimes|0\rangle\otimes|\vec p_1,s_1;\vec p_2,s_2\rangle\,.\label{final2}
\ee
The interaction Hamiltonian responsible for the transition is given by
\be
H_I=e\int d^3 x\,A_\mu(\vec x,t=0)J^\mu(\vec x,t=0)+i e_l\int d^3 x\, A_\mu(\vec x,t=0)\bar\psi \gamma^\mu\psi(\vec x,t=0)\,,
\ee
where $\psi$ is the lepton field operator. Noting that $J^\mu$ acts on $\cal{H}_{\rm QCD}$ only, and similarly
$A_\mu$ acts on ${\cal H}_\gamma$, and $\bar\psi\gamma^\mu\psi$ on ${\cal H}_l$, it is easy to see that
the matrix element of $H_I$ between our initial and final states (\ref{initial2}), (\ref{final2}) vanishes
\be
\langle f|H_I|i\rangle =0\,,
\ee
so that there is no first order transition. This brings us to consider a second order perturbation theory where the initial state first makes a transition to an intermediate state $|m\rangle$ and the intermediate state makes a transition to our final state. Inspecting $H_I$, it is clear that the intermediate state $|m\rangle$ should contain one photon quantum to
have a net non-vanishing transition by $H_I$.
For the situation like ours where the transition is allowed only at the second order perturbation theory, we have to use the corresponding Fermi's Golden rule at the second order perturbation theory,
\be
T_{i\to f}=(2\pi)\left|\sum_m{\langle f|H_I|m\rangle\langle m|H_I|i\rangle\over E_m-E_i}\right|^2 \delta(E_f-E_i)\,.
\ee
From the structure of $H_I$, we find that there are two classes of possible intermediate states

Case A) : The intermediate state involves one photon state only
\be
|m\rangle=|\alpha_m\rangle\otimes|\vec k,\epsilon_\mu^{(s)}\rangle\otimes|0\rangle\,,
\ee
and this intermediate photon decays to the final di-lepton pair.

Case B) : The intermediate state consists of one photon and the di-lepton pair,
\be
|m\rangle=|\alpha_m\rangle\otimes|\vec k,\epsilon_\mu^{(s)}\rangle\otimes|\vec p_1,s_1;\vec p_2,s_2\rangle\,,
\ee
and the intermediate photon is subsequently absorbed by the QCD plasma to leave the di-lepton pair
in the final state.

The Case A) is more intuitive from the picture of the relativistic Feynman diagram of having a virtual photon line between the QCD current and the final di-lepton pair. The Case B) in fact arises from the same Feynman diagram with a reversed time ordering where the QCD current operator appears later than the photon-lepton interaction vertex. Only after summing the two cases A) and B) in our mundane quantum mechanics treatment, we can reproduce the relativistic
result from a single Feynman diagram with a relativistic photon propagator. We will be able to check this shortly.

The quantization of the photon field $A_\mu$ in a finite volume $V$ is explained in (\ref{photonquantum}) in the Appendix 1, and we have a similar quantization of the lepton field $\psi$ as
\be
\psi(\vec x,t)={1\over V}\sum_{\vec n,s}{1\over \sqrt{2 E_{\vec p_{\vec n}}}}\left(u(\vec p_{\vec n},s)e^{-i E_{\vec p_{\vec n}}t+i\vec p_{\vec n}\cdot\vec x} a^{(s)}_{\vec p_{\vec n}}+v(\vec p_{\vec n},s) e^{+i E_{\vec p_{\vec n}}t-i\vec p_{\vec n}\cdot\vec x} b^{(s)\dagger}_{\vec p_{\vec n}}\right)\,,
\ee
where $(a^{(s)}_{\vec p_{\vec n}},b^{(s)}_{\vec p_{\vec n}})$ are annihilation operators of lepton and anti-lepton respectively, which satisfy the anti-commutation relations
\be
\{a^{(s)}_{\vec p_{\vec n}},a^{(s')\dagger}_{\vec p_{\vec n'}}\}=\{b^{(s)}_{\vec p_{\vec n}},b^{(s')\dagger}_{\vec p_{\vec n'}}\}=V\delta_{\vec n,\vec n'}\delta^{s,s'}\,.\quad
\ee
Our conventions for the Dirac spinors $u$ and $v$ are explained in (\ref{diracspinors}) in section 2.2.
The properly normalized di-lepton state in a finite volume is
\be
|\vec p_{\vec n_1},s_1;\vec p_{\vec n_2},s_2\rangle={1\over V} a^{(s_1)\dagger}_{\vec p_{\vec n_1}}b^{(s_2)\dagger}_{\vec p_{\vec n_2}}|0\rangle\,.
\ee
We are now ready to compute the matrix elements of $H_I$.

Case A) Taking the intermediate state $|m\rangle=|\alpha_m\rangle\otimes|\vec k_{\vec n},\epsilon_\mu^{(s)}\rangle\otimes|0\rangle$, we have
\be
\langle m|H_I|i\rangle={1\over\sqrt{V}}{e\over \sqrt{2|\vec k_{\vec n}|}}\epsilon_\mu^{(s)*}\int d^3 x \,e^{-i\vec k_{\vec n}\cdot\vec x} \langle\alpha_m|J^\mu(\vec x,t=0)|\alpha_i\rangle \,,
\ee
and after some algebra,
\be
\langle f|H_I|m\rangle={i\over\sqrt{V}}{e_l\over\sqrt{2|\vec k_{\vec n}|}}{1\over \sqrt{2E_{\vec p_{\vec n_1}}}}{1\over \sqrt{2E_{\vec p_{\vec n_2}}}}\epsilon_\nu^{(s)}\bar u(\vec p_{\vec n_1},s_1)\gamma^\nu v(\vec p_{\vec n_2},s_2)
\delta_{\alpha_f,\alpha_m}\delta_{\vec k_{\vec n},\vec p_{\vec n_1}+\vec p_{\vec n_2}}\,.\label{fhi}
\ee
Since the energies of the states are $E_i=\varepsilon_i$, $E_m=\varepsilon_m+|\vec k_{\vec n}|$, and $E_f=\varepsilon_f+E_{\vec p_{\vec n_1}}+E_{\vec p_{\vec n_2}}$, we have
\be
E_m-E_i=\varepsilon_m-\varepsilon_i+|\vec k_{\vec n}|=\varepsilon_f-\varepsilon_i+|\vec k_{\vec n}|=-\left(E_{\vec p_{\vec n_1}}+E_{\vec p_{\vec n_2}}\right)+|\vec p_{\vec n_1}+\vec p_{\vec n_2}|\,,
\ee
where we have used $\varepsilon_m=\varepsilon_f$ from $\delta_{\alpha_f,\alpha_m}$ in (\ref{fhi}), and
the $\delta(E_f-E_i)$ factor in the Fermi's Golden rule gives us the equality
\be
\varepsilon_f-\varepsilon_i=-\left(E_{\vec p_{\vec n_1}}+E_{\vec p_{\vec n_2}}\right)\,,
\ee
and finally we replace $\vec k_{\vec n}$ with $\vec p_{\vec n_1}+\vec p_{\vec n_2}$ due to the
$\delta_{\vec k_{\vec n},\vec p_{\vec n_1}+\vec p_{\vec n_2}}$ term in (\ref{fhi}).
Note that the sum over the intermediate states reduces to a sum over the intermediate photon polarizations only, since
$|\alpha_m\rangle=|\alpha_f\rangle$ and $\vec k_{\vec n}=\vec p_{\vec n_1}+\vec p_{\vec n_2}$.
Collecting things together, we have
\bear
\sum_m{\langle f|H_I|m\rangle\langle m|H_I|i\rangle\over E_m-E_i}\Bigg|_{{\rm A})}&=&
{i\over V}{e e_l\over (-p^0_f+|\vec p_f|)}{1\over 2|\vec p_f|}
{1\over\sqrt{2E_{\vec p_{\vec n_1}}}}{1\over\sqrt{2E_{\vec p_{\vec n_2}}}}\left(\sum_s \epsilon_\mu^{(s)*}\epsilon_\nu^{(s)}\right)\nonumber\\
&\times& \bar u(\vec p_{\vec n_1},s_1)\gamma^\nu v(\vec p_{\vec n_2},s_2)\int d^3 x\, e^{-i\vec p_f\cdot\vec x}\langle \alpha_f|J^\mu(\vec x,t=0)|\alpha_i\rangle\,,\nonumber\\\label{caseA}
\eear
where
\be
p_f^\mu=\left(E_{\vec p_{\vec n_1}}+E_{\vec p_{\vec n_2}},\vec p_{\vec n_1}+\vec p_{\vec n_2}\right)\,,
\ee
is the total energy-momentum of the di-lepton pair.

Case B) With the another class of the intermediate states
$|m\rangle=|\alpha_m\rangle\otimes|\vec k_{\vec n},\epsilon_\mu^{(s)}\rangle\otimes|\vec p_{\vec n_1},s_1;\vec p_{\vec n_2},s_2\rangle$, a similar computation gives
\be
\langle m|H_I|i\rangle={i\over\sqrt{V}}{e_l\over\sqrt{2|\vec k_{\vec n}|}}{1\over \sqrt{2E_{\vec p_{\vec n_1}}}}{1\over \sqrt{2E_{\vec p_{\vec n_2}}}}\epsilon_\nu^{(s)*}\bar u(\vec p_{\vec n_1},s_1)\gamma^\nu v(\vec p_{\vec n_2},s_2)
\delta_{\alpha_m,\alpha_i}\delta_{\vec k_{\vec n},-(\vec p_{\vec n_1}+\vec p_{\vec n_2})}\,,\label{mhi}
\ee
and
\be
\langle f|H_I|m\rangle=
{1\over\sqrt{V}}{e\over \sqrt{2|\vec k_{\vec n}|}}\epsilon_\mu^{(s)}\int d^3 x \,e^{i\vec k_{\vec n}\cdot\vec x} \langle\alpha_f|J^\mu(\vec x,t=0)|\alpha_m\rangle\,,
\ee
and the summation over intermediate states reduces to a photon polarization sum as
$|\alpha_m\rangle=|\alpha_i\rangle$ and $\vec k_{\vec n}=-(\vec p_{\vec n_1}+\vec p_{\vec n_2})$
due to the delta functions in (\ref{mhi}).
The energies of the states are $E_i=\varepsilon_i$, $E_m=\varepsilon_m+|\vec k_{\vec n}|+E_{\vec p_{\vec n_1}}+E_{\vec p_{\vec n_2}}=\varepsilon_i+|\vec p_f|+p^0_f$, and $E_f=\varepsilon_f+E_{\vec p_{\vec n_1}}+E_{\vec p_{\vec n_2}}=\varepsilon_f+p^0_f$, so that the energy denominator is
\be
E_m-E_i=p^0_f+|\vec p_f|\,,
\ee
which gives us
\bear
\sum_m{\langle f|H_I|m\rangle\langle m|H_I|i\rangle\over E_m-E_i}\Bigg|_{{\rm B})}&=&
{i\over V}{e e_l\over (p^0_f+|\vec p_f|)}{1\over 2|\vec p_f|}
{1\over\sqrt{2E_{\vec p_{\vec n_1}}}}{1\over\sqrt{2E_{\vec p_{\vec n_2}}}}\left(\sum_s \epsilon_\mu^{(s)}\epsilon_\nu^{(s)*}\right)\nonumber\\
&\times& \bar u(\vec p_{\vec n_1},s_1)\gamma^\nu v(\vec p_{\vec n_2},s_2)\int d^3 x\, e^{-i\vec p_f\cdot\vec x}\langle \alpha_f|J^\mu(\vec x,t=0)|\alpha_i\rangle\,,\nonumber\\\label{caseB}
\eear
which is almost the same with (\ref{caseA}) except the energy denominator and the complex conjugation of the polarization sum.

The polarization sums, $\sum_s \epsilon_\mu^{(s)*}\epsilon_\nu^{(s)}$ and  $\sum_s \epsilon_\mu^{(s)}\epsilon_\nu^{(s)*}$, should be replaced by a relativistic tensor,
\be
\sum_s \epsilon_\mu^{(s)*}\epsilon_\nu^{(s)}=\sum_s \epsilon_\mu^{(s)}\epsilon_\nu^{(s)*} \to \eta_{\mu\nu}\,,
\ee
in a fully relativistic quantization of the gauge field, which can be justified for example in the Gupta-Bleuler quantization that involves unphysical ghost states in a subtle way. We will here simply take it as a working recipe.

After this replacement, (\ref{caseA}) and (\ref{caseB}) differ only by the energy denominator, and the addition of the two finally gives
\bear
\sum_m{\langle f|H_I|m\rangle\langle m|H_I|i\rangle\over E_m-E_i}&=&
{i\over V}{e e_l\over p_f^2}
{1\over\sqrt{2E_{\vec p_{\vec n_1}}}}{1\over\sqrt{2E_{\vec p_{\vec n_2}}}}\bar u(\vec p_{\vec n_1},s_1)\gamma^\mu v(\vec p_{\vec n_2},s_2)\nonumber\\
&\times& \int d^3 x\, e^{-i\vec p_f\cdot\vec x}\langle \alpha_f|J_\mu(\vec x,t=0)|\alpha_i\rangle\,,\nonumber\\\label{AandB}
\eear
where $p_f^2=-(p^0_f)^2+|\vec p_f|^2$ is precisely the relativistic denominator of the photon propagator in the Feynman diagram. Therefore, the sum of the two cases A) and B) reproduces the relativistic
result.

Taking the square of (\ref{AandB}) and performing the same manipulations for $T_{i\to f}$ that we do for the case of the photon emission in the previous section noting that $E_f-E_i=\varepsilon_f-\varepsilon_i+p^0_f$,
we arrive at
 \bear
 T_{i\to f}&=&{1\over V}{e^2e_l^2\over (p_f^2)^2}{1\over 2E_{\vec p_{\vec n_1}}}{1\over  2E_{\vec p_{\vec n_2}}}\left[\bar u(\vec p_{\vec n_1},s_1)\gamma^\mu v(\vec p_{\vec n_2},s_2)\right]\left[\bar u(\vec p_{\vec n_1},s_1)\gamma^\nu v(\vec p_{\vec n_2},s_2)\right]^*G^<_{\nu\mu}(p_f)\nonumber\\
 &=&{1\over V}{e^2e_l^2\over (p_f^2)^2}{1\over 2E_{\vec p_{\vec n_1}}}{1\over  2E_{\vec p_{\vec n_2}}}(-1)\left[\bar u(\vec p_{\vec n_1},s_1)\gamma^\mu v(\vec p_{\vec n_2},s_2)\right]\left[\bar v(\vec p_{\vec n_2},s_2)\gamma^\nu u(\vec p_{\vec n_1},s_1)\right]G^<_{\nu\mu}(p_f)\,,\nonumber\\
 \eear
using the fact that $[\bar u(\vec p_{\vec n_1},s_1)\gamma^\nu v(\vec p_{\vec n_2},s_2)]^*=-[\bar v(\vec p_{\vec n_2},s_2)\gamma^\nu u(\vec p_{\vec n_1},s_1)]$.
The number of momentum pairs $(\vec p_1,\vec p_2)$ within the phase space volume $d^3 p_1d^3 p_2$ is
$V^2/(2\pi)^6 d^3 p_1d^3 p_2$, so that the total transition rate into such states is the product of $T_{i\to f}$ and $V^2/(2\pi)^6 d^3 p_1d^3 p_2$, and therefore the transition rate per unit volume and per unit phase space $d^3p_1 d^3p_2$ for a pair of lepton and anti-lepton is finally given by
\bear
{d\Gamma^{s_1,s_2}\over d^3 p_1 d^3 p_2}&=&{V\over (2\pi)^6}T_{i\to f}\\&=&{1\over (2\pi)^6}{e^2e_l^2\over (p_f^2)^2}{1\over 2E_{\vec p_1}}{1\over  2E_{\vec p_2}}(-1)\left[\bar u(\vec p_1,s_1)\gamma^\mu v(\vec p_2,s_2)\right]\left[\bar v(\vec p_2,s_2)\gamma^\nu u(\vec p_1,s_1)\right]G^<_{\nu\mu}(p_f)\,,\nonumber
\eear
after taking the infinite volume limit, which is our desired formula (\ref{dileptonappendix}).

\vfil

\end{document}